# Light Weight Implementation of Stream Ciphers for M-Commerce Applications

Mona Pourghasem, Elham Ghare Sheikhloo, Reza Ebrahimi Atani
*Department of Computer Engineering, University of Guilan, Rasht, Iran.*
m.prghsm@gmail.com, e_sheikhlou@yahoo.com, rebrahimi@guilan.ac.ir

*Abstract*

*In today's world the use of computer and telecommunications networking is essential for human life. Among these, mobile tools and devices due to availability, have found a special impact on everyone life. This feature addition to providing sample facilities such as financial transactions at any place and time has raised the Sensitivities about security of these devices. In order to provide security, numerous techniques have been proposed which due to the limitations of mobile devices; an algorithm should be taken that have the ability to function for light weight ubiquitous computing. In this paper, four eSTREAM candidates from software profile were taken into account and analyzed and implemented by using J2ME technology. Then these algorithms were implemented on a variety of mobile phones and are compared with each other in terms of execution time and finally the obtained Results are expressed.*

**Keywords:** *eSTREAM, J2ME, M-Commerce, Mobile, Light Weight, Ubiquitous Computing.*

## 1. Introduction

Expansion and infiltration of mobile tools and devices has brought many changes in the requirements of human societies. Today, mobile phones, personal digital assistants PDA and smart phones have become an indispensable tool for doing many tasks. Due to being small and relatively cheap, these devices can be used not only for make voice and video calls or send and receive text and multimedia messages, but they can also appear as a desktop computer [1].

The use of mobile devices as a desktop computer is based on the concept of ubiquitous computing that first time was proposed by Mark Weiser in 1999. Ubiquitous computing is an interactive model between human and computer, so that inside each object in our environment a computer be embedded and performed the computing and information processing operations [2]. Sought to extend this concept, the use of other tools as computers became more welcomed and considered. This issue went so far that currently 98.8 percent of the microprocessors that produced in factories are used in form of embedded microprocessor in other devices and tools and only 1.2 percent of the microprocessors are produced for use in traditional computers [3]. Also, the fast development of communication technologies, and then the wireless technologies and mobile phone networks has increased popularity of instruments and mobile devices in public. Among these, the use of mobile phone had rapid and significant growth, so that the number of mobile phone subscribers has reached from 12.4 million subscribers in 1990 to nearly 4.6 billion in 2010 [4]. The prominent and unique features of mobile phones are its perpetual access and mobility. The mobile phone always and everywhere is with the user.

In one hand mobility feature of mobile phones and in other hand doing ubiquitous computing on them, has turned the security in this tool to the sensitive and challenging subject, so far that providing security has been proposed as a new subcategory of security issues in information technology in these kinds of tools. Especially that these kinds of applications inherently are performed in insecure contexts such as wireless network or Bluetooth. For example consider the payment financial transaction. Users are willing to pay on their mobile phone only when that has the full confidence of its security.

For security, many approaches and strategies is presented such as using of encryption methods and techniques. But due to existing limitations in mobile phones that mainly is caused from shortage of energy resources, cost and low computational power, we should used algorithms that have the ability to work on this tool with considering the mentioned limitations. One of the primitives that were considered more than before is the use of Light weight cryptography algorithms and sequential algorithms which are considered a kind of lightweight. The rest of the paper is organized as in the second section the nature and the way to achieve light weight algorithms is presented. In the third part the cryptography of a light weight sequence has been proposed and the famous stream cipher algorithms are introduced. In the fourth section, the Infrastructure and technology used in the implementation of selected algorithms has been described and finally at the end, the execution time of these algorithms is compared on different cell phones and the obtained results are expressed.

## 2. Light weight cryptography

In today's world the use of decoding science is not limited only to the political and military applications, but also plays an important role in applications such as electronic commerce, mobile commerce, communication via e-mail, Internet, telecommunications, and many data exchanges. In addition, increasing the use of cryptography in devices such as credit cards, cell phones, wireless networks, remote controllers, or the alarm, has been caused that in addition to speed and security, other requirements such as cost, low power consumption and the ability to reconfigurable be raised. Thus, in recent years great efforts has been done to develop these kinds of systems in order to fix the mentioned needs. The use of cryptography in the embedded systems and systems-on- chip (SoC) is an example of the efforts made in this field. This issue puts the use of light weight cryptography algorithm in the focus, although still the most important issue in the evaluation of a cryptography algorithm is its resistance against various attacks of cipher analyzing.

Today, almost every electronic device that is used in everyday applications has security Infrastructures that provides the personal applications of that device. The security of these infrastructures is provided by using the cryptography algorithms. Design principles of light weight cryptography algorithms, is a new branch of cryptography science that are discussed as an interface of electrical engineering, cryptography and computer science. Lightweight cryptography focus is on new design or the adaptive and efficient implementation of the algorithms and existing cryptography protocols. Lightweight cryptography algorithms, besides the implementation of software for the block cipher, hash function and public key cryptography, have the ability of implementation on the hardware. In the design of light weight cryptography the balance between cost, security, and performance should be fulfilled as well. For example, in block cipher key length indicates the balance between security and cost, while in hardware implementation, number of rounds is determined the balance between cost and performance. This issue is depicted in Fig 1. Usually two of these three objectives, such as security and cost, security and performance, or cost and performance can be easily achieved; however, fulfilling all the three objectives simultaneously is very difficult. In order to fulfill the objectives mentioned in a light weight cryptography algorithm, various methods have been proposed.

In general, to achieve the light weight cryptography algorithm in very style applications such as RFID tags, we have three methods:

1- Optimizing the implementation cost of standard and safety existing algorithms.
2- A little modifying the famous and reliable existing algorithms.
3- Design a new code to low cost implementation in terms of hardware.

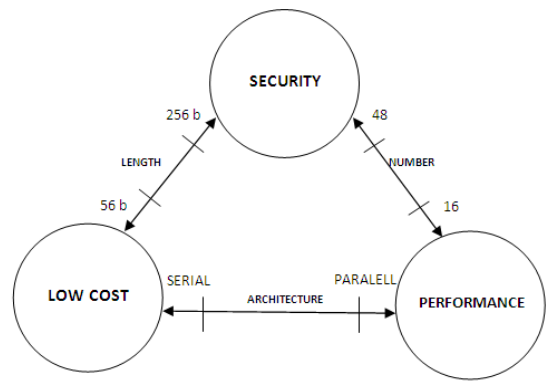

Fig. 1. Balance between performance, security, and cost in light weight cryptography [5].

The problem of first method is that most of the new block cipher algorithms are well Executable in terms of software and are not necessarily designed for hardware implementation. Today this method is suitable for block codes, because on one hand the most of different executable algorithms are exist on the PC and embedded devices in terms of software and on the other hand a cheaper silicon space, has caused that hardware implementation be performed without difficulty. If the main objective is security and low cost in devices that the both hypotheses are not considered in them, this causes that most recent block codes do not work very well on these methods. One way to bring down the implementation costs is serial processing of information and also receiving the serial data and key in input and output ports.

The second method is that to convert a famous and well known cipher to a low-cost algorithm. One of the best known algorithms in this case, is a block cipher algorithm DES (Data Encryption Standard). DES was designed for hardware implementation in the early 1970s. In [5], the Implementation In style of this algorithm has taken place with a little change as DESL. Key length of DES algorithm is not suitable for modern applications; hence the security of this algorithm was improved with acts such as key-whitening techniques [5]. Key-whitening techniques, is a method to enhance the security of block code that in it before starting the first round and after the last round, data is combined with part of a key. This combination is generally XOR operation [6].

In many cases, none of the two above methods are not appropriate to achieve an optimum design. In the third method, with design of a new algorithm which requires fewer resources we reach to the design total goals. PRESENT algorithm, is a very light weight cryptography algorithm as an example of this method has been designed and implemented [5]. According to the provided description and collective approach and competition of cryptography great scientists, about the design, analysis and implementation of a lightweight algorithms, many competitions has been held or are being held by the valid scientific societies such as NIST and ECRYPT. The purpose of these competitions is to design lightweight block, stream or hash algorithms. eSTREAM is an example of this competitions.

## 3. Lightweight Stream Cipher

eSTREAM project was initiated in late 2004 by the ECRYPT. The purpose of holding this project was to design a light and fast Stream Cipher with appropriate public acceptability. In this project, based on the expected use of a Stream Ciphers, they were divided into two sets.

Set 1: Stream Ciphers with the software applications with a high operational power.

Set 2: Stream Ciphers for hardware applications with the limited resources such as memory, gate count, and low power consumption.

A total of 34 candidates were proposed to the eSTREAM project, which at the end of competition, four oriented Stream Cipher software (set 1) were diagnosed with acceptable properties. These four Ciphers are: Rabbit, Sosemanuk, Salsa20/12, and HC-128 and also about the hardware oriented Stream Ciphers (set 2) three Ciphers namely Mickey v2, Grain v1 and Trivium were selected [6].

Stream Ciphers are an important part of private key cryptography algorithms. Using the Stream Ciphers is provided possibility of cryptography in small blocks even to the size of a character or a bit of the primary message. In the block Cipher, encode process of all data blocks, is done the same but one of the most advantages of Stream Cipher is the change of encoding functions over time. Stream Ciphers principally have less sophisticated hardware and circuits than block cipher and perform encryption operations more quickly; for this reason, in telecommunications applications that Possibility of data buffering is limited, are usually considered as the first choice. However, due to the limitation of error propagation In the Stream Ciphers, the advantage this type of code becomes more evident in applications which contain error in information propagation.

Stream Ciphers have a lot theoretical background and so far various methods and design principles have been proposed for them, and have been fully analyzed and their weaknesses have been identified. However, compared with block cipher the number of the proposed algorithm is much less and most of these algorithms because of the type of application (mobile communication) remains very specific and limited to companies or designer persons and in comparison with DES and AES standards In the block cipher, have been less available to the public. Nowadays, due to the information transmission with high speed and security on the one hand and holding projects such as eSTREAM for design efficient and secure algorithms on the other hand, Stream Cipher has been considered More than ever. Meanwhile, the selected ciphers in eSTREAM project have a special place which in the following we reviews them.

### Salsa20

Salsa20 is a family of 256-bit stream ciphers designed in 2005 and submitted to eSTREAM, the ECRYPT Stream Cipher Project. Salsa20 has progressed to the third round of eSTREAM without any changes. The 20-round stream cipher Salsa20/20 is consistently faster than AES and is recommended by the designer for typical cryptographic applications. The reduced-round ciphers Salsa20/12 and Salsa20/8 are among the fastest 256-bit stream ciphers available and are recommended for applications where speed is more important than confidence.

Salsa20 is built on a pseudorandom function based on 32-bit addition, bitwise addition (XOR) and rotation operations, which maps a 256-bit key, a 64-bit nonce (number used once), and a 64-bit stream position to a 512-bit output. This gives Salsa20 the unusual advantage that the user can efficiently seek to any position in the output stream. It offers speeds of around 4–14 cycles per byte in software on modern x86 processors[5], and reasonable hardware performance. It is not patented, and Bernstein has written several public domain implementations optimized for common architectures [7].

### Sosemanuk

Sosemanuk is a new synchronous software-oriented stream cipher, corresponding to Profile 1 of the ECRYPT call for stream cipher primitives. It uses both basic design principles from the stream cipher SNOW 2.0 [8] and transformations derived from the block cipher SERPENT [9]. For this reason, its name should refer both to SERPENT and SNOW.

Its key length is variable between 128 and 256 bits. It accommodates a 128-bit initial value. Any key length is claimed to achieve 128-bit security. The Sosemanuk cipher uses both some basic design principles from the stream cipher SNOW 2.0 and some transformations derived from the block cipher SERPENT. Sosemanuk aims at improving SNOW 2.0 both from the security and from the efficiency points of view. Most notably, it uses a faster IV-setup procedure. It also requires a reduced amount of static data, yielding better performance on several architectures [10].

### Rabbit

Rabbit is a synchronous stream cipher that was _rst presented at the Fast Software Encryption workshop in 2003 [11]. Since then, an IV-setup function has been designed [12], and additional security analysis has been completed. No cryptographical weaknesses have been revealed until now.

The Rabbit algorithm can briey be described as follows. It takes a 128-bit secret key and a 64-bit IV (if desired) as input and generates for each iteration an output block of 128 pseudo-random bits from a combination of the internal state bits. Encryption/decryption is done by XOR'ing the pseudo-random data with the plaintext/ciphertext. The size of the internal state is 513 bits divided between eight 32-bit state variables, eight 32-bit counters and one counter carry bit. The eight state variables are updated by eight coupled non-linear functions. The counters ensure a lower bound on the period length for the state variables.

Rabbit was designed to be faster than commonly used ciphers and to justify a key size of 128 bits for encrypting up to $2^{64}$ blocks of plaintext. This means that for an attacker who does not know the key, it should not be possible to distinguish up to $2^{64}$ blocks of cipher output from the output of a truly random generator, using less steps than would be required for an exhaustive key search over $2^{128}$ keys.

### HC-128

Stream cipher HC-128 is the simplified version of HC-256 [13] for 128-bit security. HC-128 is a simple, secure, software-effcient cipher and it is freely-available. HC-128 consists of two secret tables, each one with 512 32-bit elements. At each step one element of a table is updated with non-linear feedback function. All the elements of the two tables get updated every 1024 steps. At each step, one 32-bit output is generated from the non-linear output filtering function. HC-128 is suitable for the modern (and future) superscalar microprocessors. The dependency between operations in HC-128 is very small: three consecutive steps can be computed in parallel; at each step, the feedback and output functions can be computed in parallel. The high degree of parallelism allows HC-128 to run efficiently on the modern processor [14].

In order to examin the performance of these algorithms, they were implemented on different platforms such as the Intel Pentium 4 and its results have been specified. But Since the Stream Ciphers Seem like a good selection in cryptography on mobile devices With power limited resources and limited computing power it can be used to provide security mechanisms like privacy in applications that are executed on these devices, conditional on that implementation, testing and execution of these algorithms be done in the real world and can meet the needs in terms of safety, performance, cost and speed. Because of this in the paper, four selected algorithms in eSTREAM project are implemented by using J2ME technology and compared with each other on different cell phones in terms of speed. First, we will have a brief look at Implementation infrastructure of this project.

## 4. Implementation Infrastructure

When the Sun Inc. was introduced K Virtual Machine (KVM) in 1999, in the beginning, large number of software manufacturers were oblivious to it, but this virtual machine has played a valuable role in development of software for embedded systems. Before that KVM be produced and supplied, the sun Inc. was looking to realize the slogan "Write once, run everywhere" by using the Java and Java Virtual Machine (JVM). The Sun purpose was to produce the portable software for any computational platform but in the meantime a simple fact led the Sun's approach towards embedded systems and it was that most of computational platforms are into tools that have more limited resources in terms of processors and memory than the general purpose computers and therefore efforts to execute programs that was written to the Java language and fit with traditional computers, seemed pointless on this tool. Sun solved this problem by using the provided small Java edition (J2ME) and also KVM.

Despite being independent of platform, is Java's greatest strength but it cannot be ignored that Writing programs with no error and maintenance of them, by using Java is easier than C And C + +. Although Java's language has been designed on the base of C language but its designers have tried to remove many of troublemaker features of the C language. The features that sometimes have impeded the code understand and make it more difficult to maintain them. These problems sometimes led to indistinguishable programming errors. Besides, Java is an object oriented language that allows the software developers to take the most advantage from this type of language features such as Class definition, Polymorphism, Encapsulation and Inheritance in the embedded class definition of Java libraries.

With all the mentioned benefits, the size problem that is in direct contradiction with JVM and the Java standard and large libraries was existed as a major obstacle in the use of Java on embedded systems. The initial JVM of Sun Inc. required 512 Kb ROM memory which this memory was only required for the essential components of bit code execution like Bytecode interpreter, Garbage collector, Dynamic class loader, and Bytecode verifier and was not written included other components such as the program bit code. For using of JVM which had the minimum of Java special API Libraries 2 Mb ROM was needed. For address this problem, some developers which presented an embedded JVM that were smaller, faster and have more accurate unexpected collectors, but still for having Java special API libraries it was necessary to have at least 1 MB of ROM.

Sun Inc. to solve the size problem placed the Java libraries miniaturizing on the agenda as the embedded Java. As an example presented the embedded Java API library that In fact was the same Java API library with an additional class selection. So that during the program compile those classes that had not been used in program, would be removed. Though this issue reduced the program volume, but it also reduced the portability of Java and since the most important feature of Java is portability, Sun modified its approach to solving the size problem and so presented the K Virtual Machine (KVM). KVM has the advantages like flexibility, minimal use of RAM and ROM. By using KVM Java programs can be executed on 16 bit and 25 MHz processors and in systems with 128 bit ROM.

Shortly after the introduction of KVM, three separate versions of Java released. The aim of this work was announced more adaptations improvement of applications written in Java language and so a way to continue the dream of "write once, run everywhere" was opened that in embedded Java model was being faded. The difference in this version is in each one's API standard. These three versions are: Edition Enterprise (J2EE), Standard Edition (J2SE), and micro edition (J2ME). The most important factor in these three versions is subset of libraries. This means that if the application use of the library subset features for desktop and server applications is enforceable on the both platforms or if a subset of the library on a mobile phone is made on two different factories, programs written for one, also runs on the other. So the approximate purpose in provide these versions is that each library collection be a subset of the previous library. However J2ME is not an accurate subset of J2EE.

There are different subsets in J2ME library. The first group is called configuration. Current configuration is Connected Device Configuration (CDC) and Connected Limited Device Configuration (CLDC). CDC is a set of libraries for use in stationary devices, always on and has relatively permanent communication with Internet. CLDC which is a proper subset of CDC in fact is a library collection that is more suitable for use in embedded uses. CLDC is focused on the small devices that have low power Batteries or their Internet connection is slow and weak or have no connection at all. Figure 2 shows how two J2ME configuration libraries are linked with other versions of libraries. It is specified in Figure that CDC and CLDC classes are not an appropriate subset for J2SE classes. For maximum portability of applications, embedded system designers who wish to run Java applications on their systems, have to choose CDC, CLDC, or both as their configuration library.

To create configuration library framework, there are also one or more optionally profiles. These profiles are attachments to the standard libraries that are specific to a particular class in embedded systems. For example, Mobile Information Device Profile (MIDP) adds special classes of mobile phones and PDA. Practical application on any embedded system with CLDC configuration and MIDP profile can be implemented in any other platform that has the same libraries.

So with the help of KVM virtual machine and by using the J2ME many applications were produced and supplied on mobile devices, especially cell phones. The programs which have ability to run on operating systems for mobile devices such as Palm, Symbian, Windows CE, embedded Linux, and solaris. Due to the mentioned features and advantages, J2ME is one of the best options for developing applications in mobile platform. For this reason J2ME language in NetBeans 6.0 was selected for the implementation of mentioned Stream Cipher in this paper, on cell phone.

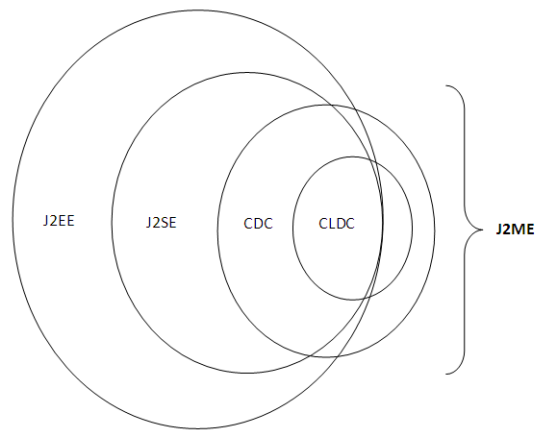

Fig. 2. Versions of the Java libraries

After the implementation of four Stream Cipher Salsa20/12, Sosemanuk, Rabbit, and HC-128 by using J2ME, this algorithm has been tested on 12 mobile phones and the execution time each of each one of them has been identified. The following points should be noted about this testing of cryptographic algorithms:
- Cell phones are made by Sony Ericsson, Samsung and Nokia.
- These phones are made between 2006 to 2010.
- Some of these phones have no operating system and some others have Symbian and Android operating systems.
- Has been given as input messages with a length of 16 to 2048 bytes to the algorithms.
- These algorithms testing have been done in natural state of mobile phone.
- The numbers presented in the table, is the execution average time of 5000 times for each algorithm.
- The presented time is in milliseconds.

# 5. Implementation Results

In the last section noted that 12 mobile phones has been selected for testing the Stream Cipher algorithms. In table 1 the information about the operating system, memory, processor, and product year of these phones has been expressed. Table 2 shows the calculation time of Stream Cipher algorithms, Salsa20/12, Sosemanuk, Rabbit, and HC-128 on mobile phones. Figure 3 and 4, respectively, is expressing line and bar graphs average algorithms execution time compared to length of the input message.

*Table1. Information of examined Mobile phones*

| Manufacture Company | Model | Operating system | Internal Memory | Processor | product year |
|---|---|---|---|---|---|
| Sony Ericsson | k800i | - | 64 MB | | 2006 |
| | P1 | Symbian v 9.1 | 16 MB MEMORY, 128 MB RAM, 256 MB ROM | 208 MHz processor، Philips Nexperia PNX4008 | 2007 |
| | vivaz | symbian series 60 v 5 | 75 MB | 720 MHz processor، PowerVR SGX GPU | 2010 |
| Samsung | s3600i | - | 30 MB | | 2008 |
| | gt-s3100 | - | 15 MB | | 2009 |
| | gt-e2152 | - | 1 MB | | 2010 |
| | Galaxy P1000 | Android V 2.2 | 16 MB MEMORY, 512 MB RAM | 1 GHz Processor ، ARM Cortex A8 | 2010 |
| Nokia | N73 | symbian 9.1 series 60 v 3 | 42 MB MEMORY, 64 MB RAM | 220 MHz processor، Dual ARM 9 | 2006 |
| | 5610 | - | 20 MB | | 2007 |
| | 3600s | - | 30 MB | | 2008 |
| | 5800 | symbian 9.4 series 60 v 5 | 81 MB MEMORY, 128 MB RAM | 434 MHz processor، ARM 11 | 2008 |
| | c6 | symbian 9.4 series 60 v 5 | 240 MB | 434 MHz processor، ARM 11 | 2010 |

*Table2. Execution time of Salsa20/12, Sosemanuk, Rabbit, and HC-128 algorithms*

| Manufacturer Company | Model | 16 | | | | 32 | | | | 64 | | | | 128 | | | | 256 | | | | 512 | | | | 1024 | | | | 2048 | | | |
|---|---|---|---|---|---|---|---|---|---|---|---|---|---|---|---|---|---|---|---|---|---|---|---|---|---|---|---|---|---|---|---|---|---|
| | | HC-128 | Rabbit | Sosemanuk | Salsa20/12 | HC-128 | Rabbit | Sosemanuk | Salsa20/12 | HC-128 | Rabbit | Sosemanuk | Salsa20/12 | HC-128 | Rabbit | Sosemanuk | Salsa20/12 | HC-128 | Rabbit | Sosemanuk | Salsa20/12 | HC-128 | Rabbit | Sosemanuk | Salsa20/12 | HC-128 | Rabbit | Sosemanuk | Salsa20/12 | HC-128 | Rabbit | Sosemanuk | Salsa20/12 |
| | | Time in milliseconds | | | | | | | | | | | | | | | | | | | | | | | | | | | | | | | |
| Sony Ericsson | k800i | 6.00 | 0.75 | 0.40 | 0.27 | 6.24 | 1.06 | 0.66 | 0.21 | 6.07 | 1.51 | 0.51 | 0.23 | 6.92 | 1.80 | 0.67 | 0.28 | 6.34 | 1.06 | 5.89 | 0.40 | 9.70 | 1.65 | 7.00 | 0.68 | 7.49 | 36.21 | 9.69 | 1.22 | 12.34 | 71.90 | 21.64 | 2.24 |
| | P1 | 7.31 | 0.71 | 1.26 | 1.33 | 6.73 | 0.53 | 1.27 | 1.66 | 7.07 | 0.70 | 1.27 | 1.21 | 6.90 | 0.82 | 1.48 | 1.89 | 7.62 | 1.25 | 1.82 | 3.63 | 8.38 | 1.30 | 2.32 | 9.43 | 10.16 | 2.35 | 3.41 | 13.54 | 10.46 | 4.36 | 5.70 | 29.35 |
| | vivaz | 1.25 | 0.14 | 0.41 | 0.12 | 1.26 | 0.15 | 0.44 | 0.16 | 1.26 | 0.17 | 0.47 | 0.11 | 1.28 | 0.21 | 0.47 | 0.13 | 1.32 | 0.28 | 0.57 | 0.11 | 1.40 | 0.52 | 0.74 | 0.25 | 1.59 | 0.87 | 1.07 | 0.35 | 1.93 | 1.68 | 2.03 | 0.81 |
| Samsung | s3600i | 8.72 | 0.65 | 1.30 | 0.89 | 8.61 | 0.57 | 1.25 | 0.86 | 8.64 | 0.71 | 1.27 | 0.85 | 8.78 | 0.82 | 1.42 | 1.53 | 9.05 | 1.07 | 1.71 | 2.82 | 9.55 | 1.64 | 2.24 | 5.38 | 10.63 | 2.87 | 3.10 | 10.53 | 12.73 | 5.23 | 5.04 | 20.64 |
| | gt-s3100 | 8.87 | 0.68 | 1.27 | 0.95 | 8.60 | 0.57 | 1.26 | 0.88 | 8.63 | 0.71 | 1.25 | 0.89 | 8.76 | 0.73 | 1.45 | 1.59 | 8.96 | 1.16 | 1.76 | 2.83 | 9.51 | 1.87 | 2.24 | 5.44 | 10.60 | 2.96 | 3.14 | 10.54 | 12.68 | 5.29 | 5.15 | 20.79 |
| | gt-e2152 | 8.71 | 0.57 | 1.32 | 0.98 | 8.44 | 0.47 | 1.32 | 0.92 | 8.43 | 0.58 | 1.24 | 0.94 | 8.65 | 0.69 | 1.43 | 1.68 | 8.87 | 0.97 | 1.84 | 3.15 | 9.52 | 1.48 | 2.45 | 6.15 | 10.72 | 2.52 | 3.59 | 12.12 | 13.42 | 4.66 | 6.04 | 24.23 |
| | Galaxy P1000 | 2.59 | 0.16 | 0.17 | 0.12 | 2.58 | 0.18 | 0.12 | 0.08 | 2.60 | 0.17 | 0.13 | 0.09 | 2.64 | 0.25 | 0.14 | 0.09 | 2.70 | 0.41 | 0.15 | 0.14 | 2.84 | 0.64 | 0.18 | 0.21 | 3.09 | 1.13 | 0.22 | 0.37 | 3.72 | 2.16 | 0.33 | 0.65 |
| Nokia | N73 | 6.66 | 0.33 | 1.21 | 0.19 | 6.56 | 0.35 | 1.20 | 0.17 | 6.63 | 0.41 | 1.22 | 0.17 | 6.62 | 0.51 | 1.39 | 0.23 | 6.87 | 0.73 | 1.74 | 0.34 | 7.25 | 1.17 | 2.26 | 0.54 | 8.60 | 2.05 | 3.27 | 0.94 | 10.83 | 4.12 | 5.55 | 1.75 |
| | 5610 | 13.70 | 0.37 | 0.35 | 0.17 | 13.70 | 0.44 | 0.34 | 0.18 | 13.80 | 0.54 | 0.34 | 0.19 | 13.98 | 0.75 | 0.39 | 0.39 | 14.40 | 1.15 | 0.47 | 0.50 | 15.15 | 1.96 | 0.59 | 0.88 | 16.70 | 3.60 | 0.84 | 1.69 | 19.75 | 6.88 | 1.38 | 3.25 |
| | 3600s | 0.18 | 0.39 | 0.34 | 0.18 | 0.19 | 0.44 | 0.33 | 0.19 | 0.19 | 0.52 | 0.33 | 0.20 | 0.29 | 0.74 | 0.38 | 0.28 | 0.49 | 1.17 | 0.46 | 0.49 | 0.89 | 1.98 | 0.59 | 0.89 | 1.70 | 3.27 | 0.83 | 1.69 | 3.31 | 7.02 | 1.37 | 3.31 |
| | 5800 | 2.01 | 0.19 | 0.57 | 0.16 | 2.01 | 0.20 | 0.52 | 0.14 | 2.03 | 0.24 | 0.52 | 0.14 | 2.05 | 0.31 | 0.59 | 0.19 | 2.13 | 0.40 | 0.73 | 0.17 | 2.28 | 0.77 | 0.94 | 0.41 | 2.59 | 1.32 | 1.36 | 0.62 | 3.29 | 2.52 | 2.29 | 1.41 |
| | c6 | 2.01 | 0.20 | 0.53 | 0.15 | 2.01 | 0.20 | 0.53 | 0.13 | 2.04 | 0.25 | 0.53 | 0.14 | 2.06 | 0.31 | 0.60 | 0.19 | 2.14 | 0.41 | 0.73 | 0.18 | 2.29 | 0.75 | 0.95 | 0.41 | 2.61 | 1.30 | 1.37 | 0.75 | 3.44 | 2.52 | 2.29 | 1.39 |
| Average | | 5.67 | 0.43 | 0.76 | 0.46 | 5.58 | 0.43 | 0.77 | 0.47 | 5.62 | 0.54 | 0.76 | 0.43 | 5.74 | 0.66 | 0.87 | 0.71 | 5.91 | 0.84 | 1.49 | 1.23 | 6.56 | 1.31 | 1.88 | 2.56 | 7.21 | 5.04 | 2.66 | 4.53 | 8.99 | 9.86 | 4.90 | 9.15 |

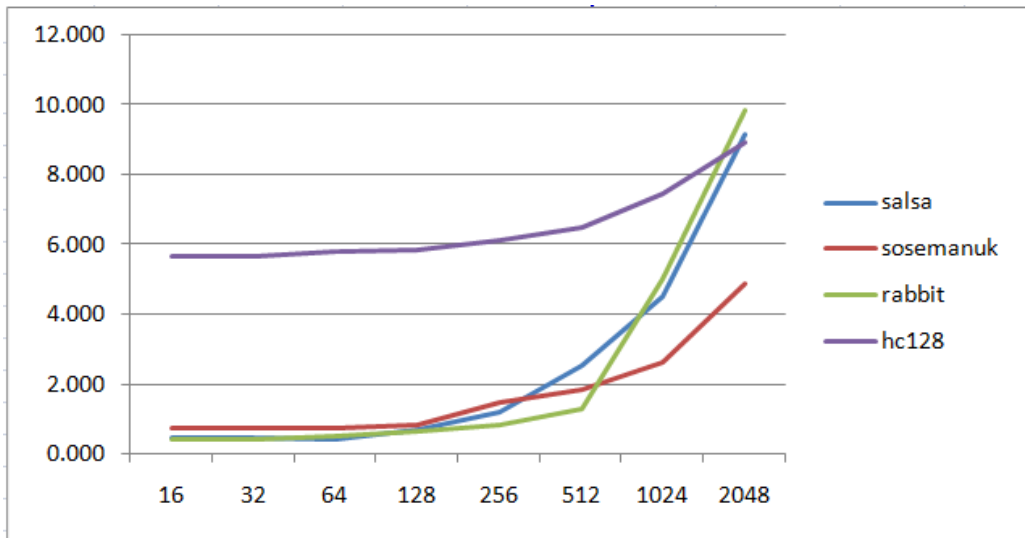

Fig3. Graph linear time - length of message 4 Stream Cipher

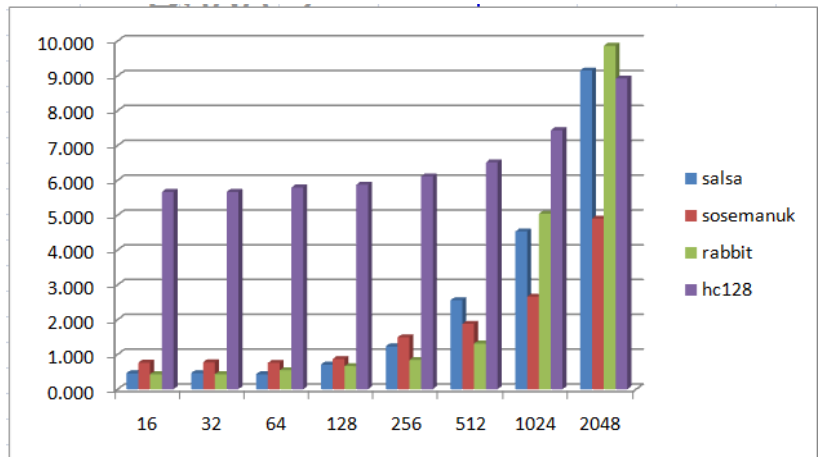

fig4. Bar graph of time - length of message 4 Cipher

The following results have been obtained from implementation, execution and testing Salsa20/12, Sosemanuk, Rabbit, and HC-128 Stream Ciphers algorithms:

- All four algorithms are implementable and executable on the mobile phone.

- Maximum execution time of algorithms Salsa20/12, Sosemanuk, Rabbit, and HC-128 for 2,048 byte input, respectively, and approximately is 9, 9, 5 and 10 milliseconds which the best time is for Sosemanuk algorithm.

- The cryptography on the 16 bit input takes less than a millisecond by using the Salsa20/12, Rabbit, and Sosemanuk algorithms, while the HC-128 algorithm is encrypted 16 bit input with a time of about 5.5 milliseconds.

- Salsa20/12 has the best performance in terms of time for messages with approximate length of 256 bytes and less, and messages with more length can be encrypted faster by Sosemanuk algorithm.

- Average execution of Salsa20/12, Sosemanuk, Rabbit, and HC-128 algorithms, respectively, are 2.44, 1.76, 2.4, and 6.4.

- By increasing the number of input bits, enhancement of cryptography time in Salsa20/12 and Rabbit algorithms is faster than Sosemanuk and HC-128 algorithms.

- On average Sosemanuk algorithm have the best and HC-128 algorithm have the worst performance in terms of execution time on mobile phones.

## Conclusions

In this paper, the necessity of providing security in mobile devices stated by using lightweight algorithms and due to the limitations existing in these tools. Then the lightweight Stream Cipher was introduced and the top four stream cryptography algorithms in eSTREAM project, namely Salsa20/12, Rabbit, HC-128, and Sosemanuk, were implemented and executed by using J2ME technology and on the 12 different mobile phones. With the execution of the algorithms, these results were obtained that all four algorithms are implemented and executed on the mobile phones. Each algorithm can be applied due to the obtained time and with respect to the size of appropriate input, for special applications, the best and worst average of execution time on the mobile phones, respectively, is belong to the Sosemanuk and HC-128 algorithms.